\documentstyle[fleqn,12pt]{article}
\begin{document}
\newcommand{\talkauthors}[1]{
 {\bf  \enskip \enskip #1}}
\newcommand{\talktitle}[1]{{\large {\bf #1}}}
\newcommand{\address}[1]{ {\it #1}}

\begin{flushright}
FIAN TD-3\\
March 1994
\end{flushright}

\vskip 30mm
\begin{center}
\talktitle{ AVERAGE MULTIPLICITIES IN GLUON AND QUARK JETS
 IN HIGHER-ORDER PERTURBATIVE QCD }

\vskip 7mm
\talkauthors{ I.M.\,Dremin\footnote{
e-mail address : dremin@td.fian.free.net}
and V.A.\,Nechitailo\footnote{e-mail address : nechit@td.fian.free.net}}

\address{\it P.N.Lebedev Physical Institute, \\ Leninsky prospect,53,
 Moscow 117924, Russia.}
\end{center}

\vskip 15mm

 \begin{abstract}
 The ratio of average multiplicities in gluon and quark jets is
 shown to become noticeably smaller in higher-order QCD compared
 to its lowest order value what improves agreement with experiment.
 QCD anomalous dimension has been calculated. It has been used to
 get energy dependence of mean multiplicities.
 \end{abstract}
 \newpage
  The ratio of average multiplicities in gluon ( $\langle n_{G}\rangle $ )
and quark ( $\langle n_{F}\rangle $ ) jets ( let us call
 it $ r \equiv \langle n_{G}\rangle / \langle n_{F}\rangle$ )
was of much debate during last years. Its value $N_c/C_F = 9/4 $
predicted in the lowest order approximation of QCD is too large to fit
experimental data. The simplest correction \cite{1,2} has reduced this
value to 2.05. Recently it has been shown \cite{3} that the exact solution
of fixed coupling QCD equations gives rise to ever smaller value
$ r = 1.84 \pm 0.02$. The above values refer to the ratio of the
number of partons in jets. In Monte Carlo computation using definite
hadronization schemes they are recalculated to hadronic ratios
observed in experiment which are about 40 percents smaller.
Using one of such models ( namely, Herwig), it has been shown in \cite{3}
 that the above partonic value of $r$ agrees quite well with experimental
\cite{4} hadronic value $1.27\pm 0.04 \pm 0.06 $.

We show here that quite similar values are obtained for the partonic ratio in
the running coupling QCD if higher order terms are taken into account.
Therefore, one concludes that this ratio is rather insensitive whether the
coupling constant is fixed or running but requires proper treatment of
energy conservation.

We have calculated also the QCD anomalous dimension in the same
approximation and have found the energy dependence of average multiplicity.
They are more sensitive to the running property of the coupling constant.

Let us remind briefly main notations. The generating functions of gluon
$G_G(y,z)$ and quark $G_F(y,z)$ jets are

\begin{equation}
G_{G}(y,z)=\sum_{n=0}^{\infty}(1+z)^{n}P_{n}^{G}(y), \,\,\,\,\,\,\,\,
G_{F}(y,z)=\sum_{n=0}^{\infty}(1+z)^{n}P_{n}^{F}(y),  \label{GGGF}
\end{equation}

\noindent where   $P^G_n(y)$ and $P^F_n(y)$  are the parton multiplicity
distributions in gluon and quark jets, correspondingly, $y = \ln(Q/Q_0)$ ,
$Q$ is the jet virtuality, $Q_0 = $const.
The normalized factorial moments $F_q$ and $\Phi_q$ are related to generating
functions as:

\begin{equation}
G_{G}=\sum_{q=0}^{\infty}\frac{z^{q}}{q!}\langle n_{G}\rangle ^{q} F_{q},
\,\,\,\,\,\,\,\,\,\,\,\,\,\,\,\,\,\,
G_{F}=\sum_{q=0}^{\infty}\frac{z^{q}}{q!}\langle n_{F}\rangle ^{q}\Phi_{q},
\label{FP}
\end{equation}
\noindent  The generating functions satisfy the following integro-differential
equations
\cite{5}

\begin{eqnarray}
\noindent G_{G}^{\prime}(y)=\int_{0}^{1}dxK_{G}^{G}(x)\gamma_{0}^{2}
[G_{G}(y+\ln x)G_{G}(y+\ln (1-x))-G_{G}(y)] \nonumber\\
\noindent +n_{f}\int_{0}^{1}dxK_{G}^{F}(x)\gamma_{0}^{2}
[G_{F}(y+\ln x)G_{F}(y+\ln (1-x))-G_{G}(y)], \label{Gprime1}
\end{eqnarray}
\begin{equation}
G_{F}^{\prime}(y)=\int_{0}^{1}dxK_{F}^{G}(x)\gamma_{0}^{2}
[G_{G}(y+\ln x)G_{F}(y+\ln (1-x))-G_{F}(y)], \label{Gprime2}
\end{equation}

\noindent where $G^{\prime}(y)=dG/dy,\,\, \gamma_{0}^{2}=2N_c\alpha_{S}/\pi,
\,\, \alpha_{S} = 2\pi/(11-\frac{2}{3}n_f)y$ is the running coupling constant,
$n_f$ is the number of active flavours of quarks.

The kernels of equations are given by QCD as :

\begin{equation}
K_{G}^{G}(x)=\frac{1}{x}-(1-x)[2-x(1-x)], \label{kgg}
\end{equation}
\begin{equation}
K_{G}^{F}(x)=\frac{1}{4N_{c}}[x^{2}+(1-x)^{2}], \label{kgf}
\end{equation}
\begin{equation}
K_{F}^{G}(x)=\frac{C_{F}}{N_{c}}\left (\frac{1}{x} -1+\frac{x}{2}\right ),
\label{kfg}
\end{equation}
\noindent where $N_{c}=3$ is the number of colours and
$C_{F}=( N_c^2 - 1)/2N_c = 4/3 $.

In solving eqs.(\ref{Gprime1}),(\ref{Gprime2}) we follow the method
developed in \cite{6,7} and expand the generating functions in Taylor
series at large $y$. We leave
aside all terms with second and higher derivatives here because their
contribution to average multiplicities is of higher order in coupling
constant as can be easily checked. In that case,
eqs.(\ref{Gprime1}),(\ref{Gprime2}) become

\begin{eqnarray}
(\ln G_{G}(y))^{ \prime}=\int^{y}_{0}dy'\gamma_0^2(y')[G_G(y')-1]
\nonumber \\
-h_1\gamma_0^2[G_G(y)-1] + h_2\gamma_0^2G_G^{ \prime}(y)
\nonumber \\
+ \frac{n_f}{12N_c}\gamma_0^2 \left\{
2\left[\frac{G_F^{ 2}(y)}{G_G(y)} - 1 \right]
- h_4\frac{G_F^{ 2}(y)}{G_G(y)}\left( \ln G_F(y) \right)^\prime \right\} ,
\label{cutoff1}
\end{eqnarray}
\begin{eqnarray}
(\ln G_{F}(y))^{ \prime}=\frac{C_F}{N_c}\left\{
\int^{y}_{0}dy'\gamma_0^2(y')[G_G(y')-1]
\right. \nonumber \\
\left.
+ \gamma_0^2 \left[ - \frac{3}{4}( G_G(y) -1 ) + \frac{7}{8}G_G^{ \prime}(y)
- h_3( \ln G_F(y) )^\prime G_G(y) \right] \right\} ,
\label{cutoff2}
\end{eqnarray}

\noindent where
$h_1 = 11/24 , h_2=(67-6\pi^2)/36 , h_3=(4\pi^2-15)/24 , h_4=13/3 $.

To calculate average multiplicities, one
should keep only the terms with
$q = 0$ and 1 in (\ref{FP}). Normalization condition is
$F_{0}=F_{1}=\Phi_{0}=\Phi_{1}=1$ . After differentiation, one gets
from (\ref{cutoff1}),(\ref{cutoff2})

\begin{eqnarray}
\langle n_{G}\rangle^{\prime\prime} = \gamma_0^2\left\{
(1+4Bh_1\gamma_0^2)\langle n_{G}\rangle
-2h_1\langle n_{G}\rangle^{\prime}
+ h_2\langle n_{G}\rangle^{\prime\prime}
\right.
\nonumber \\
+\frac{n_f}{12N_c}\left[
2 ( 2\langle n_{F}\rangle^{\prime}-\langle n_{G}\rangle^{\prime} )
-h_4\langle n_{F}\rangle^{\prime\prime}
\right.
\nonumber \\
\left.
\left. -\, 4B\gamma_0^2( 2\langle n_{F}\rangle-\langle n_{G}\rangle )
\right] \right\} ,
\label{nG}
\end{eqnarray}
\begin{eqnarray}
\langle n_{F}\rangle^{\prime\prime} = \frac{C_F}{N_c}\gamma_0^2 \left\{
(1+\frac{3}{2}B\gamma_0^2) \langle n_{G}\rangle
- \frac{3}{4}\langle n_{G}\rangle^{\prime}
 + \frac{7}{8}\langle n_{G}\rangle^{\prime\prime}
- h_3\langle n_{F}\rangle^{\prime\prime}  \right\} ,
\label{nF}
\end{eqnarray}

\noindent where $\gamma^\prime_0 = - B\gamma_0^3 , \,\,
B \equiv ( \frac{11}{3}N_c-\frac{2}{3}n_f)/8N_c$
has been substituted. The QCD anomalous dimension $\gamma(y)$
is defined as

\begin{equation}
\langle n_{G}\rangle = \exp(\int^y\gamma(y')dy') .
     \label{anomDim}
\end{equation}

\noindent We represent $\gamma$ and $r$ expanding them in $\gamma_0$ as

\begin{equation}
\gamma = \gamma_0 ( 1 - a_1 \gamma_0 - a_2 \gamma_0^2 ) + O(\gamma_0^4 ) ,
\label{decomp}
\end{equation}
\begin{equation}
r = \frac{N_c}{C_F}( 1 - r_1\gamma_0 - r_2\gamma_0^2 ) + O(\gamma_0^3 ) .
\end{equation}

Since $\gamma_0^{ \prime} \sim \gamma_0^3 $, we get $r^\prime = 0$ up to
terms $ O(\gamma_0^3)$. However, such terms are taken into account in
the formulae below.
Up to correction terms of order $O(\gamma_0^2)$ one gets

\begin{equation}
\frac{\langle n_{G}\rangle^\prime}{\langle n_{G}\rangle} =
\gamma_0\left[ 1 - a_1\gamma_0 - a_2\gamma_0^2 \right] \ ,
\label{ngp}
\end{equation}
\begin{equation}
\frac{\langle n_{G}\rangle^{\prime \prime}}{\langle n_{G}\rangle} =
\gamma_0^2\left[ 1 - (2a_1 + B)\gamma_0 -
( 2a_2-a_1^2-2a_1B)\gamma_0^2 \right] \ ,
\label{ngpp}
\end{equation}
\begin{equation}
\frac{\langle n_{F}\rangle^{\prime}}{\langle n_{G}\rangle} =
\frac{C_F}{N_c}\gamma_0\left[ 1 + (r_1 - a_1)\gamma_0 +
(r_2 + r_1^2 - a_2 - a_1r_1 - B r_1)\gamma_0^2 \right] \ ,
\label{nfp}
\end{equation}
\begin{eqnarray}
\frac{\langle n_{F}\rangle^{\prime \prime}}{\langle n_{G}\rangle} =
\frac{C_F}{N_c}\gamma_0^2\left[ 1 + (r_1 - B - 2a_1)\gamma_0 +
(r_2 + r_1^2 + a_1^2 -
\right.\nonumber \\
\left.
 3B r_1 - 2a_1r_1 - 2a_2 + 2B a_1)\gamma_0^2\right] \ .
\label{nfpp}
\end{eqnarray}

We insert (\ref{ngp})-(\ref{nfpp}) into (\ref{nG}),(\ref{nF}) and keep
again the terms of order $O(\gamma^2_0)$. Therefrom we obtain after
simple algebraic manipulations

\begin{equation}
 a_1 = h_1+\frac{n_f}{12N_c^3} - \frac{B}{2} ,
\end{equation}
\begin{equation}
 r_1 = 2\left( h_1 + \frac{n_f}{12N_c^3} \right) - \frac{3}{4} ,
\end{equation}
\begin{eqnarray}
a_2 = - \frac{1}{2}\left\{ h_2 - a_1^2
 - 2a_1\left(B-[h_1+\frac{n_f}{12N_c^3}(1+ 4C_F N_c) ] \right)
 \right. \nonumber \\
 \left.
 - \frac{n_f}{12N_c}\frac{C_F}{N_c}[h_4+3-4B] + 4B[h_1+\frac{n_f}{12N_c^3}]
\right\} ,
\end{eqnarray}
\begin{eqnarray}
r_2 = \frac{r_1}{6}\left( \frac{25}{8} -
\frac{3}{4}\frac{n_f}{N_c} - \frac{C_F}{N_c}\frac{n_f}{N_c} \right)
 + \frac{7}{8} - h_2 - \frac{C_F}{N_c}h_3
 + \frac{n_f}{12N_c}\frac{C_F}{N_c} h_4 .     \label{r2}
\end{eqnarray}

Thus, we have obtained the QCD anomalous dimension and the ratio of average
multiplicities in gluon and quark jets as functions of the QCD coupling
constant $\gamma_0$ and of the number of active flavours $n_f$ in
higher-order perturbative QCD.
The higher order corrections in $r$ are negative and its value becomes smaller.
The above dependences are very mild that agrees with conclusions of \cite{3}.
Besides, here we can insert the running coupling $\gamma_0$ in
(\ref{decomp}) and
show dependence of $\gamma$ on $Q$ directly ( see Fig.1 ). To do that, we use
( as in \cite{3} ) the parameter $Q_0 =0.65\Lambda_{\rm\overline{MS}}$ with
$\Lambda_{\rm\overline{MS}}=175 {\rm MeV}$ for $n_f = 5$ in proportion
$63:100:130$ for $n_f = 5:4:3$, correspondingly.
Integration of these values of $\gamma$ gives rise according to (\ref{anomDim})
to the following energy dependence of the average multiplicity

\begin{equation}
\langle n_{G}\rangle = y^{ \frac{ a_1}{ 2B}}
\exp\left(\sqrt{\frac{2}{B}y} + \frac{a_2}{\sqrt{2B^3y}}\right)
\label{nGy}
\end{equation}

\noindent or in terms of $\gamma_0$ one obtains

\begin{equation}
\langle n_{G}\rangle = y^{ \frac{ a_1}{ 2B}}
\exp[2\gamma_0(y)(1+a_2\gamma_0^2(y))y] \ .
\label{nGgamma}
\end{equation}

\noindent First term in the exponent corresponds to DLA-approximation
with running coupling constant. There is an important difference between
the fixed and running coupling constant impact on the energy behaviour
of average multiplicities. Even though the fixed coupling case provides
faster asymptotical increase of average multiplicity, in the energy range
of interest up to $y \sim 3y_{Z^0}$ the mean multiplicity for fixed
coupling increases slower than for running coupling. ( Here $y_{Z^0}=6.67$
is the scale at which the value of the fixed coupling has been calculated).
It is shown in Fig.2.

The physical origin of the effect is easy to understand because rather
small value of the fixed coupling has been used throughout the whole
evolution of the cascade while in real situation it should become larger
and larger at later stages what produces more particles in case of
running coupling at lower energies. However, the multiplicity is enlarged
if more flavours become active. It compensates somehow the lower values
of the coupling constant at high virtualities what should be taken into
account  when comparison with experiment is done.

The preexponential term in expressions (\ref{nGy}),(\ref{nGgamma})
is also well
known ( see, e.g., \cite{5,9,10} ) and stems from MLLA-approximation
\footnote{ Calculations in the framework of this approximation see,
for example, in refs.\cite{11,12}. }.
The next-to-next-to leading correction to the exponent calculated by
us diminishes it and the energy increase of the average multiplicity
becomes somewhat slower ( see Fig.3 ).

In Table \ref{Tab1} we show the values of the ratio $r$ for various $n_f$ at
$Q = 91$ GeV ( $Z^0$ mass ). They are very close to those obtained for
the fixed coupling QCD \cite{3} and are about 20 percents smaller than the
prediction of the lowest order approximation.
We would like to point out that the $\gamma_0^2$-correction to $r$
calculated above is several times larger than its value found in \cite{10}
where the complicated diagram calculations have been done.
The result of \cite{10} is reproduced by the first term in (\ref{r2}).
Other terms correspond to the contribution of second derivatives in
(\ref{nG}),(\ref{nF}).
For convenience, the numerical values of $a_i, r_i$ are shown in the
Table also.

Let us remind that one should combine these results with models for the
conversion of partons into hadrons ( hadronization ) to get an access to
experimental data. That is why, in ref.\cite{3}, the results of Monte
Carlo calculation according to Herwig model \cite{8} have been used.
They show that at $Z^0$-mass the ratio $r$ at parton level exceeds its
value at hadron level by about 40 percents. Being taken into account,
this fact leads to agreement between theory and experiment \cite{8}.
Since our results are very similar to those of \cite{3} we do not
foresee any contradiction between them and experimental data.

To conclude, we have obtained the QCD anomalous dimension $\gamma$ and
the ratio of average multiplicities in gluon and quark jets in higher-order
perturbative QCD with running coupling constant. The value
of the ratio agrees rather well with the one previously obtained
in fixed coupling QCD \cite{3} by exact
solution of the equations. Thus, it shows that the running property of the
coupling constant is not as important for the ratio as proper treatment
of higher-order terms including conservation laws but it is more
noticeable in the energy dependence of mean multiplicity.
Once again, it demonstrates that the
values of $r$ are much smaller than their lowest order counterparts
and agree with experiment if the hadronization is properly taken into
account. Nevertheless, one may still worry that the hadronization stage
gives rise to the most important factor reducing parton ratios to their
hadron values accessible in experiment. However, to get it in QCD one
must solve the confinement problem that is out of our scope nowadays.
\\

{\Large {\bf \noindent Acknowledgements}}
\\

We are grateful to R.Hwa for valuable comments which
help correct some formulae and improve the content.

This work was supported in part by Russian Fund for Fundamental
Studies ( grant 93-02-3815 ), by NATO grant CRG-930025 and
by International Science Foundation grant M5V000.

\newpage
\vspace{2cm}
\begin{center}
Figure Captions
\end{center}

\begin{description}
\item{Fig.\,1.} The dependence of the anomalous dimension $\gamma$ on $Q$.
Solid lines are for $\gamma$ and dashed lines are for $\gamma_0$ .
The numbers of active flavours are indicated near lines in all figures.

\item{Fig.\,2.} The $y$-dependence of the average multiplicity.
Solid lines are for running coupling and dashed lines are for fixed coupling.
The arrow marks the $y_{Z^0}$ location.

\item{Fig.\,3.} The $y$-dependence of the average multiplicity for the running
coupling. Solid lines are for next-to-next-to leading approximation considered
in the present paper and dashed lines are for MLLA.
\end{description}

\newpage
\vspace*{1cm}
\begin{table}
\hspace*{4cm}\caption{} \vspace{2mm} \label{Tab1}
\begin{tabular}{|c|c|c|c|c|c|c|}
\hline
$\quad n_f \quad$  &  $\ \quad r\quad \ $ &  $r_1$
& $r_2$ & $\gamma_0$ & $ a_1 $ & $ a_2 $\\
\hline
3 &  1.84 & 0.185 & 0.426 &  0.473 & 0.280 & - 0.379\\
\hline
4  & 1.80 & 0.191 & 0.468 & 0.481 & 0.297 & - 0.339\\
\hline
5 &  1.77 &  0.198 & 0.510 & 0.484 & 0.314 & - 0.301\\
\hline
\end{tabular}
\end{table}

\end{document}